\documentclass[12pt]{article}

\usepackage{xcolor}

\usepackage[T1]{fontenc}
\usepackage{ae,aecompl}

\usepackage{natbib}

\usepackage{graphicx}	
\DeclareGraphicsExtensions{ .pdf, .png}

\usepackage{hyperref}

\begin{document}
\title{An empirical, Bayesian approach to modelling the impact of weather on crop yield: maize in the US}


\author{
\parbox{\linewidth}{
Raphael Shirley,$^{1,2,3}$ 
Edward Pope,$^{4}$
Myles Bartlett,$^{5}$
Seb Oliver,$^{1,2}$
Novi Quadrianto,$^{5}$
Peter Hurley,$^{6}$
Steven Duivenvoorden,$^{1,2}$
Phil Rooney,$^{6}$
Adam B. Barrett,$^{2}$
Chris Kent,$^{4}$
and James Bacon.$^{4}$
}
}
\maketitle

\parbox{\linewidth}{
{\scriptsize
$^{1}$Astronomy Centre, Department of Mathematical and Physical Sciences, University of Sussex, UK\\
$^{2}$Data Intensive Science Centre, Department of Mathematical and Physical Sciences, University of Sussex, UK\\
$^{3}$Instituto de Astrof\'{i}sica de Canarias, E-38205 La Laguna, Tenerife, Spain;\\ Universidad de La Laguna, Dpto. Astrof\'{i}sica, E-38206 La Laguna, Tenerife, Spain\\
$^{4}$Met Office Hadley Centre, Exeter, UK\\
$^{5}$Predictive Analytics Lab, Department of Engineering and Informatics, University of Sussex, UK\\
$^{6}$Data Javelin Ltd, Data science consultancy, Brighton, UK
}
}



\begin{abstract}
We apply an empirical, data-driven approach for describing crop yield as a function of monthly temperature and precipitation by employing generative probabilistic models with parameters determined through Bayesian inference. Our approach is applied to state-scale maize yield and meteorological data for the US Corn Belt from 1981 to 2014 as an exemplar, but would be readily transferable to other crops, locations and spatial scales. Experimentation with a number of models shows that maize growth rates can be characterised by a two-dimensional Gaussian function of temperature and precipitation with monthly contributions accumulated over the growing period. This approach accounts for non-linear growth responses to the individual meteorological variables, and allows for interactions between them. Our models correctly identify that temperature and precipitation have the largest impact on yield in the six months prior to the harvest, in agreement with the typical growing season for US maize (April to September). Maximal growth rates occur for monthly mean temperature 18-19$^\circ$C, corresponding to a daily maximum temperature of 24-25$^\circ$C (in broad agreement with previous work) and monthly total precipitation 115 mm. Our approach also provides a self-consistent way of investigating climate change impacts on current US maize varieties in the absence of adaptation measures. Keeping precipitation and growing area fixed, a temperature increase of $2^\circ$C, relative to 1981-2014, results in the mean yield decreasing by 8\%, while the yield variance increases by a factor of around 3.  We thus provide a flexible, data-driven framework for exploring the impacts of natural climate variability and climate change on globally significant crops based on their observed behaviour. In concert with other approaches, this can help inform the development of adaptation strategies that will ensure food security under a changing climate.
\end{abstract}




\section{Introduction}

Establishing the climate risk to the global production of individual crops, and how that might change in the future, is an essential requirement for building a resilient and robust food system that ensures food security for all \citep[][]{fao:2002}. Decision-makers can then use this information to guide the development of suitable adaptation and mitigation strategies across different time frames. This requires the characterisation of the relationship between meteorological and food production variations.

There is a growing consensus that a range of methods are needed to accurately assess climate impacts on crop yield \citep[e.g.][]{lobell:2017, tigchelaar:2018, snyder:2018}. Exploring different model formulations and assumptions (e.g. a multi-model ensemble) provides a way of assessing key uncertainties and biases in our understanding of crop-climate interactions. In turn, this can help evaluate our confidence in the direction and magnitude of climate change impacts on food production. Broadly, there are two complementary approaches to this:  physiological processes-based models; or data-driven, statistical models. Physiological models are generally built upon an experiment-based understanding of the generic crop. Data-driven, statistical models can be developed when there is sufficient empirical yield data. Each approach can thus be developed and applied when the requirements for the other are not met.

Examples of statistical approaches include non-parametric models which are formulated in terms of meteorological variables rather than underlying physiological processes or critical thresholds. These models are calibrated using historical data and have demonstrated the ability to capture broad influences of weather on crop yield \citep[e.g.][]{schlenker:2009, lobell:2010, welch:2010}. This approach differs to parameterised models, calibrated by field experiments, which do account for specific thresholds in quantifying the response to temperature \citep[e.g.][]{cutforth:1990, yin:1995, wang:1998, yan:1999, streck:2007, zhou:2018} or precipitation \citep[e.g.][]{cakir:2004, ge:2012, lobell:2013, carter:2016, song:2019}. Observations have also been used to constrain parameters of more complex process-based models \citep[e.g.][]{iizumi:2009, tao:2009}. A physically motivated, but empirical, data-driven approach would complement both process-based crop models and existing statistical approaches. This formulation would allow models to be developed without extensive field trials, and with a greater range of validity.

Our approach exploits Bayesian inference to derive an empirical and non-linear ``growth response function'' that maps temperature and rainfall conditions to crop growth. This work is part of a trend to apply advanced statistics and machine learning methods to climatological and agricultural data sets. For example, \citet{you:2017} demonstrated the application of deep learning and Gaussian processes to predict yield based on remote imaging. 

Bayesian inference, is an established approach for inferring the posterior values of model parameters, based on prior assumptions and new data. The advantage of Bayesian inference is it allows robust computation of errors, which is especially critical when the aim is to model the influence of predicted climate data which are themselves subject to large uncertainties. More generally, Bayesian inference is frequently applied to problems of model parameter estimation with noisy data in other fields such as astronomy \citep{Hurley:2016}. In this paper, we are determining generative probabilistic models, which have a greater ability to accurately capture uncertainty than the more common discriminative models in machine learning.  

The aim here is to present a simple and robust model which captures the impact of mean temperature and precipitation changes on mean yield. Developing a flexible method for investigating the influence of present-day natural climate on yield allows us to explore the direction and scale of climate change impacts in the absence of adaptation. We apply these methods to model the response of US maize to temperature and precipitation, which has been widely studied \citep[e.g.][]{schlenker:2009, hatfield:2011, roberts:2012, lobell:2013, sanchez:2014, hatfield:2015, Partridge:2019}. We aim to first demonstrate an ability to capture key aspects of present-day maize yield variability in the US, and secondly to explore the implications of climate change for the current maize varieties in the absence of adaptation. 

The paper is structured as follows: section~\ref{sec:data} describes the yield and meteorological data, followed by section~\ref{sec:models} which presents the models, demonstrating their strengths and weaknesses. Section~\ref{sec:impacts} applies the model to an ensemble of climate projections as a first step to predict the influence of climate change on yield for present day maize varieties. Finally, section~\ref{sec:conclusions} summarises the work and our main conclusions.

\section{Data}
\label{sec:data}

Numerous studies have demonstrated the importance of temperature in maize development \citep[e.g.][]{cross:1972, coelho:1980, daughtry:1984, cutforth:1990, bonhomme:1994}, with quantities such as Growing Degree Days (GDD) offering better predictions of phenological changes and yield than calendar days after planting. Water stress is also associated maize yield reductions \citep[e.g.][and references therein]{cakir:2004, ge:2012, lobell:2013, carter:2016, song:2019}, suggesting that any model should incorporate the effects of both temperature and precipitation.

We proceed assuming that climate variability is a major driver of observed yield anomalies, and do not attempt to separate direct physiological influences from impacts resulting from air quality or pests \citep{Gornall:2010}. We use annual maize yield and monthly temperature and rainfall data aggregated to the state scale. This minimises the impact of local variations in both meteorology and planting date \citep[e.g.][]{schlenker:2009, lobell:2010}. For future work, the model outputs are compatible with the large-scale atmospheric circulation patterns that can be reliably simulated by global climate models. The relatively low data requirements, compared to daily data, also support model flexibility and computationally efficiency. Constraining the model parameter posteriors depends on the sampling of temperature/precipitation space, the data can only constrain features which are present in the data. We therefore draw observations from across the US Corn Belt, covering Indiana, Illinois, Iowa, Ohio, Minnesota and Nebraska to maximise the sampling while being confident the regions use broadly similar agricultural techniques and have comparable climatologies.

The spatial distribution of maize cultivation (both irrigated and rained) was extracted from the MIRCA2000 dataset \citep{Portmann:2010} and used to derive area-weighted climate variables. Monthly mean temperature and monthly total precipitation are based on CRU TS3.1/3.21 \citep{Harris:2014} and GPCC v6 \citep{Schneider:2014} and extracted for 1981-2014 from within the WFDEI dataset \citep{Weedon:2014}. An overview of the data is shown in Table~\ref{table:data}. 

\begin{table*}
\centering
\caption{
Standard deviations of monthly temperature and precipitation relative to the corresponding monthly climatological mean across all months, and the standard deviation of the yearly yield anomalies. Yield information is available from 1960, and the gridded WFDEI data extends back to 1980. Values shown here are from 1980 to present. Standard deviations in the yield are for anomalies from the rolling five year median yield.}

\label{table:data}
\vskip 0.15in
\begin{tabular}{llll}
State  & 
Monthly $\sigma_{\Delta T}$   & 
Monthly $\sigma_{\Delta P}$   & 
Yearly $\sigma_{\Delta Y}$   \\

& [$^\circ$C] & [mm] & [t ha$^{-1}$]\\
\hline
Indiana & 1.93 & 35.25 & 1.68 \\ 
Illinois & 2.02 & 36.39 & 1.78 \\ 
Ohio & 1.87 & 31.05 & 1.67 \\ 
Nebraska & 2.21 & 30.90 & 1.95 \\ 
Iowa & 2.23 & 33.55 & 1.92 \\ 
Minnesota & 2.39 & 29.06 & 2.06 \\ 
\end{tabular}
\end{table*}

In all regions there has been a long-term increase in maize yield since 1960, on top of which year-to-year and multi-year variations are evident. To remove the long-term trend and decadal-scale variability, which can be driven by climate and non-climate factors \citep[e.g.][]{hawkins:2013, ray:2015}, we use yield `anomalies'. Removing the estimated long term trend gives an estimate of the yield anomaly relative to what was expected that year, which retains the units of tonnes per hectare (t/ha). We compare a range of methods for calculating yield anomalies: a) anomalies relative to the centred 5-year rolling median yield in each region which is subtracted from the time series; b) anomalies relative to the 5-year rolling mean; c) fractional anomalies relative to the 5-year rolling mean (which are unit-less); d) anomalies relative to the least squares linear trend across all regions. \citet{finger:2010}, applied robust detrending techniques, which are designed to perform regression in the presence of outliers. Such an approach could be used here to perform the linear de-trending; however, it would still rely on the assumption of a linear productivity increase. Removing outliers beyond 10\% fractional yield anomaly influenced the linear fit parameters by less than 10\% and had a negligible impact on parameter posteriors (changes to median values of less than 5~\% of the variance).

The total yield anomaly in year $j$, $\Delta Y_j$, is the difference between the yield for year $j$, $Y_j$, and the rolling five-year median or mean $\widetilde{Y}_j = (Y_{j-2},Y_{j-1},Y_{j},Y_{j+1},Y_{j+2} )$. A feature of using the rolling 5-year median anomalies is that, on average, one in five anomalies will be exactly zero. For this reason we compare all four forms of yield anomaly. The choice of de-trending has some influence on the posterior parameter values and, therefore, the model performance as described in section~\ref{sec:validation}. The temperature and precipitation anomalies for each month are calculated by subtracting the corresponding monthly climatological mean for 1981-2014. However, the main model presented uses actual temperature and precipitation values and not anomalies.

\begin{figure}
\centering
\includegraphics[width=0.8\textwidth]{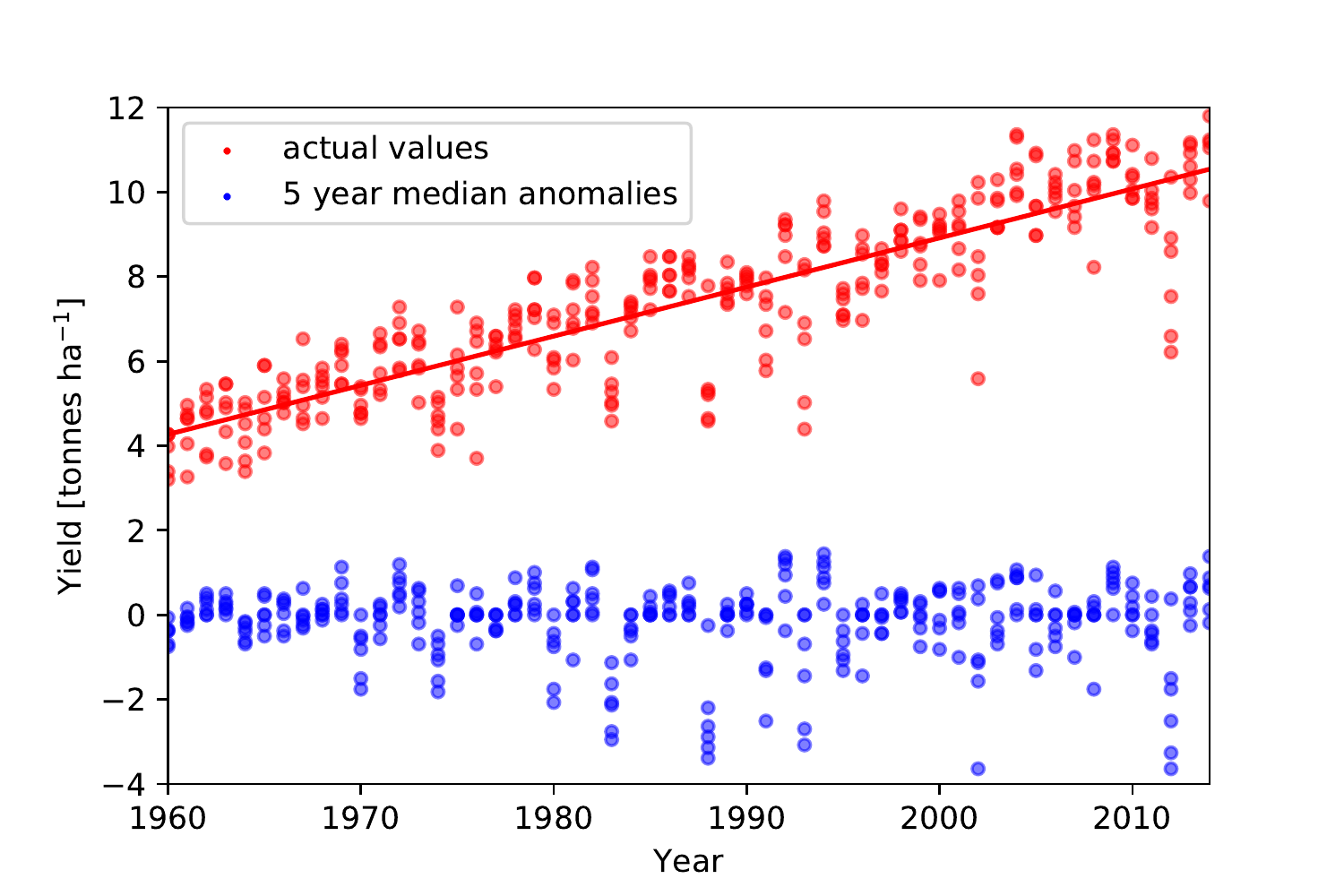}
\caption{\label{fig:yield_anomalies} The annual yield for all six US states used in this work (Indiana, Illinois, Ohio, Nebraska, Iowa and Minnesota), alongside the anomalies derived from subtracting the running five-year median for the yield data set from 1960 to 2014. We also show the linear fit which is used to scale all the anomalies to the year 2014, noting that this is different to the moving 5-year median.}
\end{figure}

\begin{figure}
\centering
\includegraphics[width=0.8\textwidth]{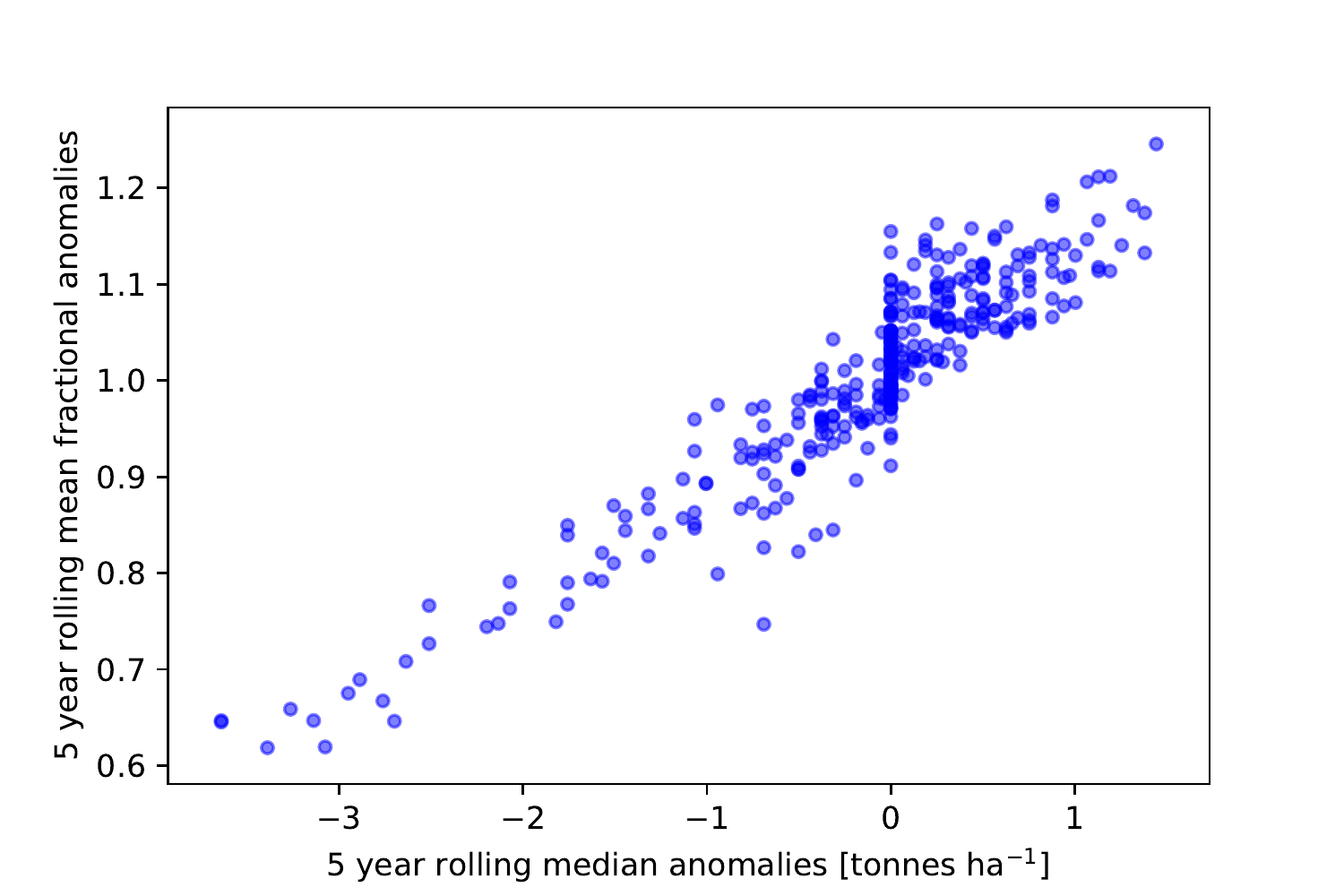}
\caption{\label{fig:median_vs_frac_anoms_noline} Comparison between the 5-year median anomalies and the 5-year mean fractional anomaly. A feature of the 5-year median anomaly is that roughly one fifth of all anomalies are exactly zero, which introduces non-Gaussian behaviour. In contrast, a benefit of using the fractional difference to the 5-year mean is that we might expect anomalies to increase in proportion to the mean yield. The preferred model is the one which is most highly correlated with temperature and precipitation.
}
\end{figure}

The relationship between the yield and the temperature and precipitation anomalies forms the basis of our linear models. These provide a useful comparator for more complex models that relate yield to temperature and precipitation directly. Table~\ref{table:data} summarises the variation around typical values for monthly temperatures and precipitation and annual maize yields. Figure~\ref{fig:yield_anomalies} illustrates how these two measures are related, while Figure~\ref{fig:median_vs_frac_anoms_noline} shows the relation between the 5-year median anomaly and the 5-year fractional mean anomaly. Later, we compare what impact the choice of target yield has on the performance of models.

Previous analysis of these data by \cite{kent:2017} revealed that yield reductions greater than 10\% are strongly associated with mean temperatures during June, July and August exceeding 23 $^\circ$C, combined with total precipitation less than 240 mm. In short, warmer and drier than normal conditions were associated with reduced maize yield in the US Cornbelt. There was also evidence that excess precipitation during the same period could predict yield reductions; however, this relationship was more tentative. 

\begin{figure}
\centering
\includegraphics[width=0.8\textwidth]{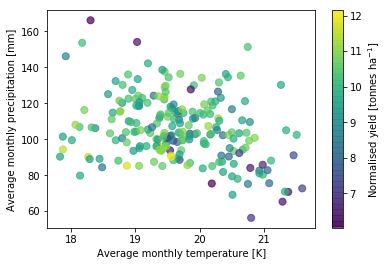}
\caption{\label{fig:real_yield_responses} Mean maize yield, normalised to the 2007 linear trend value, as a function of mean monthly temperature and precipitation for the months April to September between 1980 and 2007 for Indiana, Illinois, Ohio, Nebraska, Iowa and Minnesota.}
\end{figure}

Figure~\ref{fig:real_yield_responses} shows the maize yield time series, normalised to 2007 linear trend levels, as a function of yearly average temperature and precipitation. 
Even at this level of temporal coarse-graining, (i.e. averaging temperature and precipitation anomalies over six months), there is evidence that warmer and drier conditions are associated with significantly reduced maize yield, and the basic features that any model should try to capture are evident. 

\section{Methods and models}
\label{sec:models}

It is well known that both temperature and water availability play influential roles in determining maize growth and yield. Temperature, in particular, affects the rate of phenological development \citep[e.g.][]{cross:1972, coelho:1980, daughtry:1984, cutforth:1990, bonhomme:1994}, with a range of studies demonstrating that growth and yield have a non-linear dependence on both daily temperature and accumulated thermal sums \citep[e.g.][]{cutforth:1990, streck:2007, schlenker:2009, lobell:2013, zhou:2018}. The non-linear response often incorporates cardinal temperatures which describe minimum, optimal and maximum thresholds for a particular crop \citep[e.g.][]{yin:1995, wang:1998, zhou:2018}. While many models make use of standard cardinal temperatures \citep[e.g.][]{yin:1995}, this work uses the observed data to estimate the optimal growing temperature.

Water stress is also associated with maize yield reductions \citep[e.g.][and references therein]{cakir:2004, ge:2012, lobell:2013, carter:2016, song:2019}, and can be affected by precipitation and temperature as well as soil and land management strategies. The joint dependence on precipitation and temperature can be understood in terms of the plant's demand for water, which is related to vapour pressure deficit (VPD) between the saturated plant leaf and the ambient air \citep[e.g.][]{roberts:2012, lobell:2013}. Higher VPD tends to occur on warmer days with lower humidity, promoting higher transpiration rates from the plant. The plant may respond to reduce water loss by reducing stomatal conductance, but this can inhibit metabolic activity and carbon assimilation, potentially resulting in yield failure \citep[e.g.][]{song:2010, ge:2012, lobell:2013, song:2019}. The empirical response of US maize yield to precipitation is weaker than for temperature \citep[e.g.][]{lobell:2013}, but water availability remains an important consideration and we, therefore, include precipitation in the models outlined below.

Building on the previous research outlined above, the models developed here explore yield dependence on monthly temperature and rainfall accumulated during the growing season, making use of both linear and non-linear models trained on monthly weather observations. This approach allows us to explore the simultaneous influence of monthly temperature and precipitation variations on US maize, as well as the effect of interactions between these two variables. Using monthly resolution will also allow the model to applied in situations where the only data available on on that time scale such as is common from, for instance, satellite imaging derived data. As we add complexity to these models, we can capture more features in the data, but the model may also be subject to parameter degeneracies and over-fitting, given the limited volume of data we have chosen to fit against. The classes of model investigated here are:

\begin{itemize}
\item Linear models, predicting yield using a sum over the growing season of the monthly contributions to growth based on a linear independent function of observed temperature and precipitation anomalies.
\item Gaussian process regression, predicting yield using correlations between each monthly temperature and precipitation and their correlation with the target yield.
\item Two-dimensional Gaussian model, predicting yield using a sum over the growing season of the monthly contributions to growth based on a two-dimensional Gaussian function (i.e. non-linear) of observed temperature and precipitation.
\end{itemize}

The linear model was used to determine which months were most closely correlated with yield. Using this as a basis, the parameters of the linear and two-dimensional Gaussian models were initially inferred using least squares minimisation, and subsequently using Bayesian inference to investigate the full posterior on the model parameters. In principle, it is possible to use the regression coefficients to extract information about the crop's response to temperature and precipitation, hereafter referred to as the growth response function. However, using the Gaussian models described below, we are able to make a more direct estimate of the growth response function.

Gaussian process regression was used as a baseline to measure the capacity of meteorological data to predict the yield. This is because it is a general form of regression which does not make any assumptions about the `true' model as parameterised models must.

\subsection{Growth as function of temperature and precipitation}

After initial investigations using the linear model and Gaussian process regression to identify those months where climatology is correlated with the yield, we extended the approach to develop a more physically-motivated model which allows for non-linear responses to both temperature and precipitation. As a heuristic, we assume there is an optimum temperature and precipitation (subject to other variables being held constant, e.g. solar radiation, soil type and CO$_{2}$ concentration) away from which the plant's growth rate declines \citep[c.f.][]{cutforth:1990, yin:1995, wang:1998, hatfield:2015, korres:2016, tigchelaar:2018}. Here we model the monthly contribution to yield as a time-independent two-dimensional Gaussian function of mean monthly temperature and precipitation. This allows us to implicitly incorporate the effects of exposure to cold and heat as well as insufficient and excess precipitation, without explicitly needing to derive critical thresholds (e.g. cardinal temperatures) from the data. Since the function is time-independent it represents the crop's typical response to growing conditions averaged across the entire growing season. For an individual state, 30 years of monthly data is insufficient to fully sample the $T$, $P$ plane around the peak of the Gaussian; to combat these data limitations we combine information from the different states. The model is described in full in section~\ref{sec:2d_descrip_appendix} and in the notebooks which execute the code and are available on GitHub.

The approach developed here explains the results in figure~\ref{fig:real_yield_responses} through developing a more general and continuous description of the crop's response to temperature and precipitation, while maintaining consistency with the threshold-based approach demonstrated in \cite{kent:2017}. This approach shares some similarities with \cite{snyder:2018} who developed an emulator of process-based crop models based on Agricultural Model Intercomparison and Improvement Project (AgMIP) Coordinated Climate-Crop Modeling Project (C3MP) data.

Freely fitting the Gaussian parameters for each month during the calendar year led to large unconstrained posteriors on the parameters. This is predominantly a consequence of the volume of data relative to the number of parameters, and means that further assumptions are needed to reduce the number of free parameters. One option is to assume the same functional shape for each month, while allowing the critical temperatures to change during different growth stages \citep[e.g.][]{hatfield:2011, sanchez:2014}. This assumption reduces the model parameters by a factor of twelve. Based on the growing season for US maize, we also restricted equation~\ref{eqn:yield_sum} to sum over months April to September (months 4-9). This was determined using the linear response model which showed that months 1-3 had a correlation with yield that was consistent with zero. We experimented with various priors and settled on wide Gaussian priors around the means of the measurements, checking that the priors did not have a large influence on the posterior.

In the next two subsections we describe the models, while their performance is discussed in \ref{sec:validation}.

\subsection{Linear model}
\label{sec:appendix_linear}

The first model investigated here used multiple linear regression to predict yield anomalies as a function of monthly mean temperature and precipitation anomalies, with their contributions to maize growth summed over the growing season. The physical interpretation of this approach is that the regression coefficients capture the crop's underlying response to climate variables such that the yield anomaly is determined by that year's temperature and precipitation throughout the growing season. This method was used to empirically determine the months to include in training and prediction. The tightest correlation was between the months of April and September -- as expected under the assumption that the months leading up to and including the harvest are critical for determining yield.

In this linear model and the later two dimensional Gaussian model, the yield anomaly in year $j$ is given by the sum of the monthly mean growth rates, $\dot{y}_{i,j}$, for each month $i$ during the growing season:

\begin{equation}
\Delta Y_j = \sum_{i=1}^{N} \Delta y_{i,j} = \sum_{i=1}^{N} \dot{y}_{i,j}\Delta t_i
\end{equation}

Where $\Delta t_i$ is the duration of the monthly interval, and the growth rate is some function of the monthly temperature anomaly $\Delta T_i$ and precipitation anomaly $\Delta P_i$, defined as the difference of the month $i$ measurement relative to the 30-year mean for that month.

Under these assumptions, the growth rate can be expressed as
\begin{equation}
\dot{y}_i = f(T_i, P_i) \approx s_t(\Delta T_i) + s_p(\Delta P_i)
\end{equation}
The approximation assumes that the function is slowly changing on the scale of the temperature and precipitation anomalies such that a Taylor expansion to first order is sufficient and that the variations with temperature and precipitation are independent of each other. In this limit, the regression coefficients, $s_t$ and $s_p$, will be related to the gradient of the growth response and are, themselves, functions of temperature and rainfall anomalies.

\subsection{Gaussian process regression}
\label{sec:GP_descrip_appendix}
A Gaussian process is a non-parametric distribution over an infinite collection of random variables such that any finite subset constitutes a multivariate normal distribution \citep{rasmussen:2005}. Such models are completely specified by their second-order statistics, namely the mean and covariance, though most often assume the former to be zero everywhere and rely entirely on the latter to evaluate the predictive distribution. The covariance function measures similarity in the input space; a common choice for this is the squared exponential covariance function which we employ here. 

Gaussian processes serve as a general procedure for predicting non-linear behaviour; as such it provides a baseline against which to compare the predictive power of any physically-motivated parameterised model. Here, we use the Gaussian process model to provide a baseline for the power of temperature and precipitation to predict yield. As such, this helps test the sufficiency of linear models and our suggested generative model outlined below.

\subsection{Two-dimensional Gaussian yield response function}
\label{sec:2d_descrip_appendix}

This section outlines the assumptions and mathematical formulation that underpin the bivariate Gaussian yield response function. Validation metrics shown in section~\ref{sec:validation} demonstrate that yield predictions made by the bivariate Gaussian outperform those for both the linear and the Gaussian process models. For that reason, this discussion goes into more depth than the previous section.

Being non-linearly dependent on temperature and precipitation, the bivariate Gaussian model shares some common features with \cite{cutforth:1990, streck:2007, yin:1995, zhou:2018}. However, there are several major differences to previous work: firstly, the model incorporates both temperature and precipitation, allowing for potential interactions between their impacts on growth; secondly, the non-linear function is Gaussian rather than a Beta function \citep[e.g.][]{yin:1995, streck:2007}, which avoids the need to estimate explicit minimum and maximum thresholds for temperature and precipitation since the growth rate tends to zero for large deviations from optimal growing conditions; thirdly, the contributions to maize growth are calculated and summed on monthly intervals, rather than daily as for Growing Degree Day models \citep[e.g.][]{zhou:2018}.

The mathematical structure of the model is as follows. The $k$ data points for each year that we are fitting with the model are 
\begin{equation}
X_k = (Y_k, \mathbf T_k,  \mathbf P_k)
\end{equation}
where $\mathbf T_k$ and $\mathbf P_k$ are vectors with the temperature and precipitation values, respectively, for each month of the year. Under this model the monthly yield response (increase in final yield due to that month's conditions) is
\begin{equation}
\dot{y}(T,P) =  \exp\left(-\frac{1}{2} ({\mathbf x}-{\mathbf \mu})^{\mathbf T} {\mathbf \Sigma}^{-1}  ({\mathbf x}-{\mathbf \mu})\right)
\end{equation}
where
\begin{equation}
\mu =  \left[ {\begin{array}{c}
   \mu_T \\
   \mu_P \\
  \end{array} } \right]
\end{equation}
and
\begin{equation}
\label{eqn:2d_gaussian_sigma}
    \mathbf\Sigma = \left[ {\begin{array}{cc}
    \sigma_T^2 & \rho \sigma_T \sigma_P \\
    \rho \sigma_T \sigma_P  & \sigma_P^2 \\
    \end{array} } \right]
\end{equation}
The predicted yield is then
\begin{equation}
\label{eqn:yield_sum}
Y = \sum_i n_i \dot{y}_i
\end{equation}
where, $n_i$ is the normalisation for the $\it{i}$th month. There are also many variations on this general structure; for instance instead of using all twelve preceding months we can use six, as with the linear regression model. We can also force the parameters of the Gaussian shape to be constant across time while allowing the normalisation to vary. This latter option assumes that favourable conditions are a constant over the growing period but will allow for months closer to harvest to have a more significant impact on the yield. The final model we present assumes that the normalisation is fixed across the six months prior to harvest.

\begin{figure}
\centering
\includegraphics[width=0.8\textwidth]{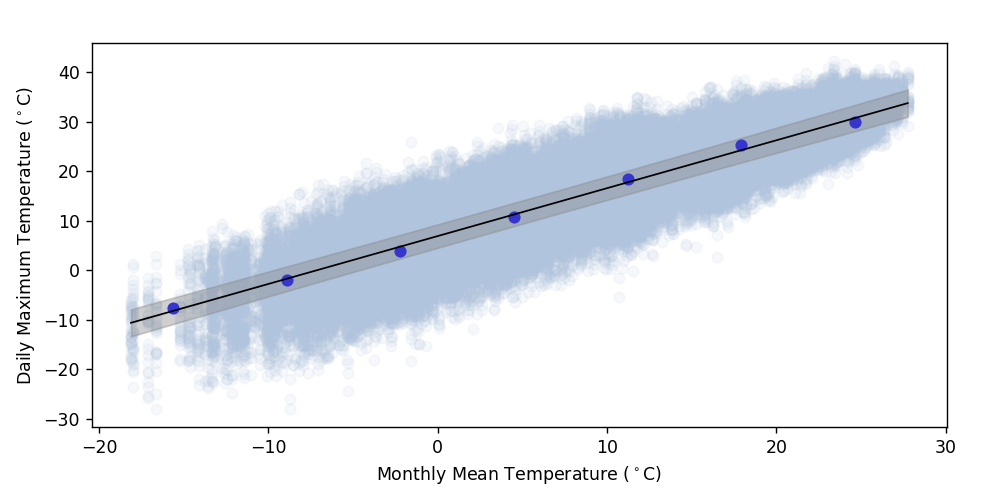}
\caption{\label{fig:daily_vs_monthly} Dependence of mean daily maximum temperature on monthly mean temperature for Indiana, Illinois, Ohio, Nebraska, Iowa, Minnesota. The Pearson correlation coefficient is 0.998, the gradient is 0.967 $\pm$ 0.024 and the intercept 6.863 $\pm$ 0.34. As a result, a monthly mean temperature of 18--19~$^\circ$C translates to an optimal daily maximum of 24--25~$^\circ$C. }
\end{figure}

Mathematically, the likelihood (of one data point) is the probability of the data given the model,
\begin{equation}
P\{ Y, \mathbf{T}, \mathbf{P} | M\left(\mathbf{n}, \mu_T, \sigma_{T}, \mu_P,\sigma_{P},\rho\right) \}
\end{equation}
Assuming that we know $\mathbf{T}$ and $\mathbf{P}$, the probability of one yield data point is 
\begin{equation}
P\{ Y | M(\mathbf{n}, \mu_T, \sigma_{T}, \mu_P,\sigma_{P},\rho),  \mathbf{T}, \mathbf{P} \}
\end{equation}
In this case, the likelihood is 
\begin{equation}
\textrm{ln} \mathcal{L} = \sum_k \textrm{ln} \left( \mathcal{N}(Y - \sum_i n_i \dot{y}_i, \sigma ) \right)
\end{equation}
Although the model is a function of monthly mean temperatures, figure~\ref{fig:daily_vs_monthly} shows how the parameters can be related to daily maximum temperatures through their strong correlation with monthly mean temperatures. The likelihood and the prior jointly determine the posterior. The Stan language and its Python wrapper PyStan are used here to sample the posterior model parameter space \citep{carpenter:2017}.

\subsection{Model Validation}
\label{sec:validation}

\begin{figure}
\centering
\includegraphics[width=0.8\textwidth]{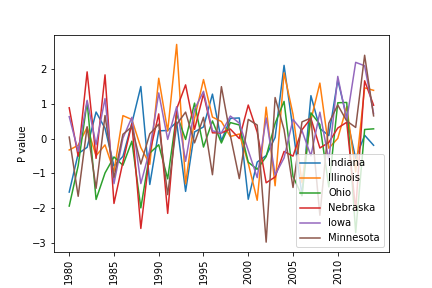}
\caption{\label{fig:pvalues_all} The Bayesian p-values for each year's yield measurement to demonstrate the performance of the model. This is equivalent to the distance of the true value to the predicted value in terms of model variance.If the model is capturing behaviour these values will be normally distributed with unity variance.}
\end{figure}

This section outlines a selection of methods and procedures for assessing the predictive performance of the statistical models described in the previous section. We evaluate the robustness of our model using a suite of cross-validation schemes. The first method uses Bayesian p-values to quantify how unlikely the observed yield is, given the model. If the observed data is unlikely given the proposed model, the differences between them will be large compared to the uncertainty in the prediction. Figure~\ref{fig:pvalues_all} shows these p-values for each year and state. This method does not split the data set into training and testing sets so over-fitting is a possibility. We, therefore, investigate a number of more rigorous approaches below.

\begin{figure}
\centering
\includegraphics[width=0.8\textwidth]{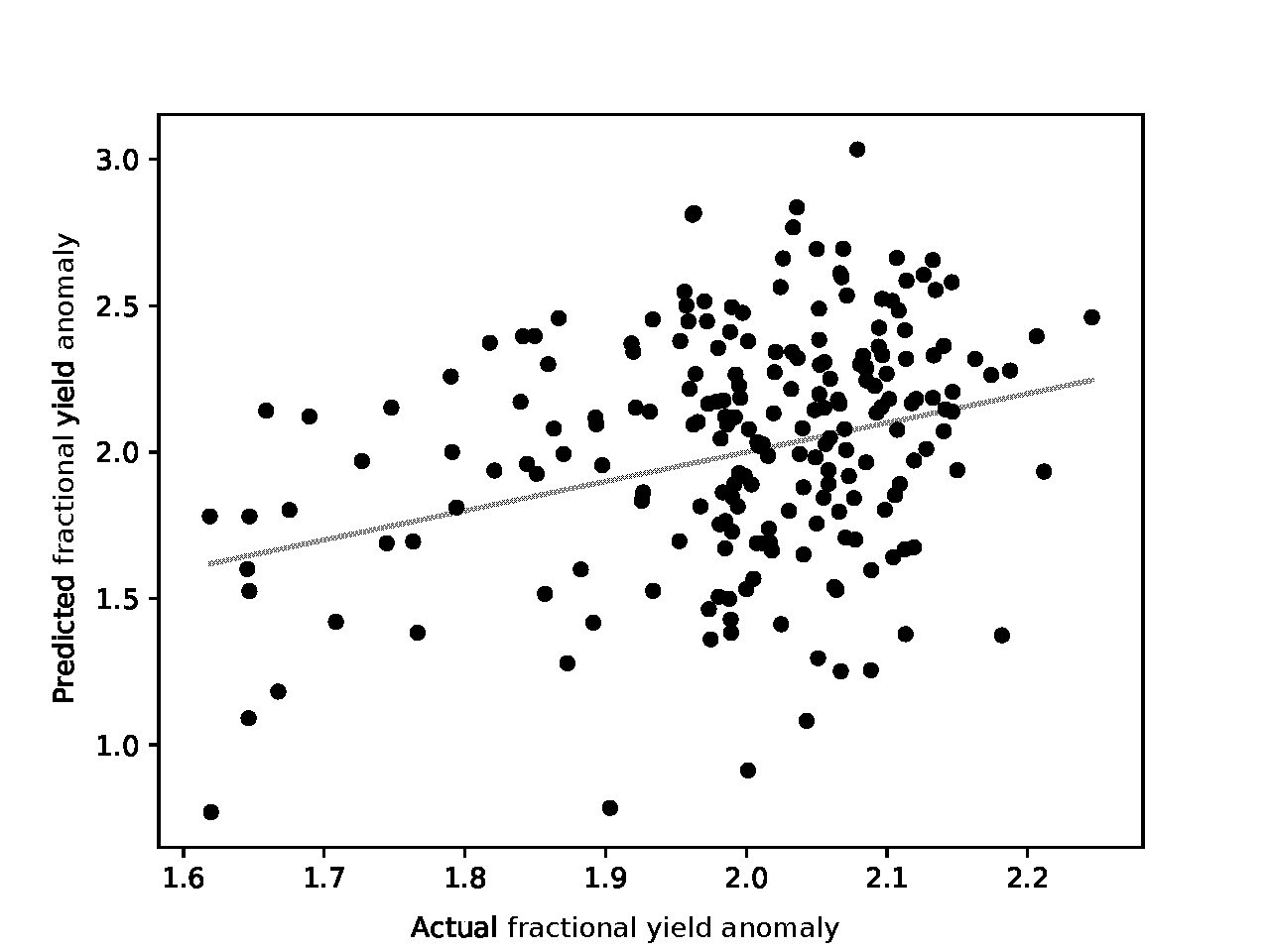}
\caption{\label{fig:cv_plot} Predicted versus observed maize fractional yield anomalies for the six US states during 1980-2014. The grey line shows equality between predicted and observed values. The predicted values are based on the posterior median which was used for the computation of the classical $R^2$. The full posterior captures the full distribution of predicted values for the full parameter sample and is measured by the Bayesian $R^2$. Predictions were generated under LOO cross-validation.}
\end{figure}

Since our model is time-stationary by construction, we first consider methods which do not aim to remove temporal dependencies. The most straightforward of these is to randomly split the data such that eighty percent of the years are used to infer the regression coefficients, while the remaining twenty percent are withheld to assess the predictions. This stochastic procedure is repeated with ten different seeds to obtain an average. In leave-one-out (LOO) validation, we test on each year in turn, retaining the data from all other years for training. The average annual predicted versus observed fractional yield for each year under this regime is shown in Figure~\ref{fig:cv_plot}. We also include a variation of this in which we exclude adjacent yields when inferring the regression parameters, following \citep{iizumi:2018}, to reduce potential temporal correlation.

A yet stricter approach is to follow a rolling-origin validation procedure, such that only observations prior to the one currently being tested on are available for training: in each iteration we advance forward a year and accumulate an additional training sample. We begin by employing the standard root-mean-square error (RMSE) and classical $R^2$, indicating the degree of variance captured by the predictions, as quality measures of our cross-validated estimates. Table~\ref{table:cv_results_clas_bayes} summarizes the results for the three models obtained with the various validation schemes using the five year median anomaly. The RMSE values for each of the three schemes are all comparable with Gaussian process regression marginally performing best according to most metrics. The negative classical $R^2$ values show that all of the point predictions do not predict year to year variations if we ignore errors and factors that are not modelled in the measured yield. This means that the model point predictions can be used to predict \emph{mean} yield changes resulting from climate changes but not to accurately predict yearly yields.

The classical $R^2$ values fail to make full or appropriate use of the model posteriors. Indeed, \citet{Gelman:2019} argued that classical $R^2$ is also not an appropriate metric of model performance in the context of Bayesian inference due to the possibility of it lying outside the [0,1] interval. Instead, they suggest a new metric, ``Bayesian $R^2$", which lies with the [0,1] interval by construction and involves using draws from the posterior rather than the mean or median values used by classical $R^2$. The output of a Bayesian model is a probability distribution for every measurement, such that using a point prediction from the posterior median parameter values is arbitrary. The Bayesian $R^2$ is designed to provide a measure of the variance in the data accounted for by the model, and is defined as the predicted variance divided by predicted variance plus error variance:

\begin{equation}
\label{eq}
\textrm{Bayesian } R^2_s = \frac{V^N_{n=1} y_n^{ s} }{V^N_{n=1} y_n^{ s}  + \textrm{var}^s_{\mathrm{res}}}
\end{equation}
where $V^N_{n=1} y_n^{s}$ is the variance of the predicted values for the draw from the posterior, $s$, and $\textrm{var}^s_{\mathrm{res}}$ is the expected residual variance. This equation therefore describes the proportion of variance explained for a given draw from the posterior sample. We therefore use these Bayesian $R^2$ values for the posteriors, computed here as the median of twenty draws from the posterior. 

Using the Bayesian $R^2$ measure, the bivariate Gaussian model performs the best out of the suite of similar models considered here. The results also imply that the 5-year mean fractional anomaly is the best captured of all the yield metrics. We, therefore, recommend the use of the bivariate Gaussian monthly growth model with the 5-year mean fractional anomaly values, and the remainder of the paper focuses on this model.

\begin{table}
\small
\centering
\caption{Comparing Root Mean Square Error (RMSE), classical $R^2$, and Bayesian $R^2$ values calculated from 20 draws from the posterior for the median five-year anomaly values. Both RMSE and classical $R^2$ require a point prediction to be calculated which is not naturally generated by a Bayesian model. In order to calculate it for the purpose of the metric we use the posterior median parameter values. Due to the limitations of the classical metrics we use Bayesian $R^2$ values to compare model performance, which is designed to be a measure of the fraction of variance explained by the model. It is difficult to interpret classical $R^2$ values that lie outside the [0,1] interval and, since Bayesian modelling is not designed to provide a point estimate, it is considered somewhat arbitrary to investigate the posterior median.}
\label{table:cv_results_clas_bayes}
\vskip 0.15in
\begin{tabular}{lrrr}
\hline
 & \multicolumn{3}{c}{5 year median anomaly}  \\
Validation Method   & RMSE   & Classical $R^2$  & Bayesian $R^2$   \\ \hline

\multicolumn{4}{c}{Linear Regression}    \\ \hline
Random Split        & 0.91   & 0.03   &  0.20 \\
LOO                 & 0.85   & -2.73  &  0.04 \\
Modified LOO        & 0.83   & -2.45  &  0.20 \\
Rolling-origin      & 1.05   & -5.82  &  0.05 \\ \hline
\multicolumn{4}{c}{Gaussian Processes}   \\ \hline
Random Split        & 1.00   & -0.01  &  0.03 \\
LOO                 & 0.86   & -2.43  &  0.04 \\
Modified LOO        & 0.85   & -2.38  &  0.09 \\
Rolling-origin      & 0.92   & -4.03  &  0.04 \\ \hline
\multicolumn{4}{c}{Bivariate Gaussian}   \\ 
\hline
Random Split        & 1.09   & -4.34  &  0.40 \\
LOO                 & 1.05   & -4.05  &  0.34 \\
Modified LOO        & 1.01   & -4.53  &  0.40 \\
Rolling-origin      & 1.34   & -3.56  &  0.35 \\ \hline
\end{tabular}
\end{table}

\begin{table}
\small
\centering
\caption{Bayesian $R^2$ values for the three models, obtained using the four different cross-validation schemes. The classical $R^2$ is computed from the posterior median parameter values (which may not be mutually consistent). Bayesian $R^2$ values are the mean from 20 draws from the posterior. All anomalies are computed from the five-year window centered on the yield value, except the linearly detrended yields which use the anomaly from the all state linear trend from least squares fitting. The highest values are found for the bivariate Gaussian model using five-year mean fractional anomalies.}
\label{table:cv_results}
\vskip 0.15in
\begin{tabular}{l llll}
\hline
Validation Method     & 5-year median         & 5-year mean          & 5-year mean fraction  & Linear detrending   \\ \hline

\multicolumn{5}{c}{Linear Regression}   \\ \hline
Random Split          & 0.20                  & 0.07                 & 0.08                  &  0.07  \\
LOO                   & 0.04                  & 0.18                 & 0.19                  &  0.06  \\
Modified LOO          & 0.20                  & 0.21                 & 0.21                  &  0.24  \\
Rolling-origin        & 0.05                  & 0.05                 & 0.05                  &  0.06  \\ \hline
\multicolumn{5}{c}{Gaussian Processes}   \\ \hline
Random Split          & 0.03                  & 0.05                 & 0.04                  &  0.04 \\
LOO                   & 0.04                  & 0.05                 & 0.05                  &  0.06 \\
Modified LOO          & 0.09                  & 0.10                 & 0.10                  &  0.10 \\
Rolling-origin        & 0.04                  & 0.05                 & 0.05                  &  0.05 \\ \hline
\multicolumn{5}{c}{Bivariate Gaussian} \\ \hline
Random Split          & 0.40                  & 0.37                 & 0.49                  &  0.40 \\
LOO                   & 0.34                  & 0.50                 & 0.49                  &  0.36 \\
Modified LOO          & 0.40                  & 0.40                 & 0.51                  &  0.43 \\
Rolling-origin        & 0.35                  & 0.33                 & 0.50                  &  0.36 \\ \hline
\end{tabular}
\end{table}

\section{Results}
\label{sec:results}
\begin{figure}
\centering
\includegraphics[width=0.8\textwidth]{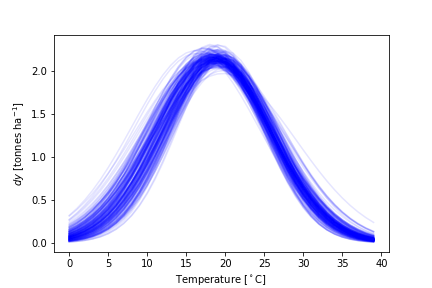}
\caption{\label{fig:2d_Gauss_temp_post_sample_growth_curve} A sample of posterior Gaussian parameter values presented as monthly growth expressed as a function of temperature for the bivariate Gaussian model at $P = 100$mm. Note there is greater uncertainty for low temperatures, which occur rarely during the maize growing season in the US. Each line corresponds to a set of parameter values drawn from the posterior.}
\end{figure}

\begin{figure}
\centering
\includegraphics[width=0.8\textwidth]{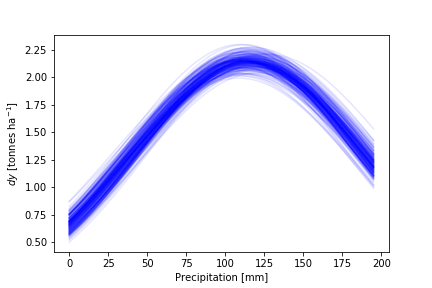}
\caption{\label{fig:2d_Gauss_precip_post_sample_growth_curve} A sample of posterior Gaussian parameter values presented as monthly growth expressed as a function of precipitation value at $T = 20$~$^\circ$C. Note there is greater uncertainty for high precipitation totals, which occur rarely during the maize growing season. Each line corresponds to a set of parameter values drawn from the posterior.}
\end{figure}

\begin{figure}
\centering
\includegraphics[width=0.8\textwidth]{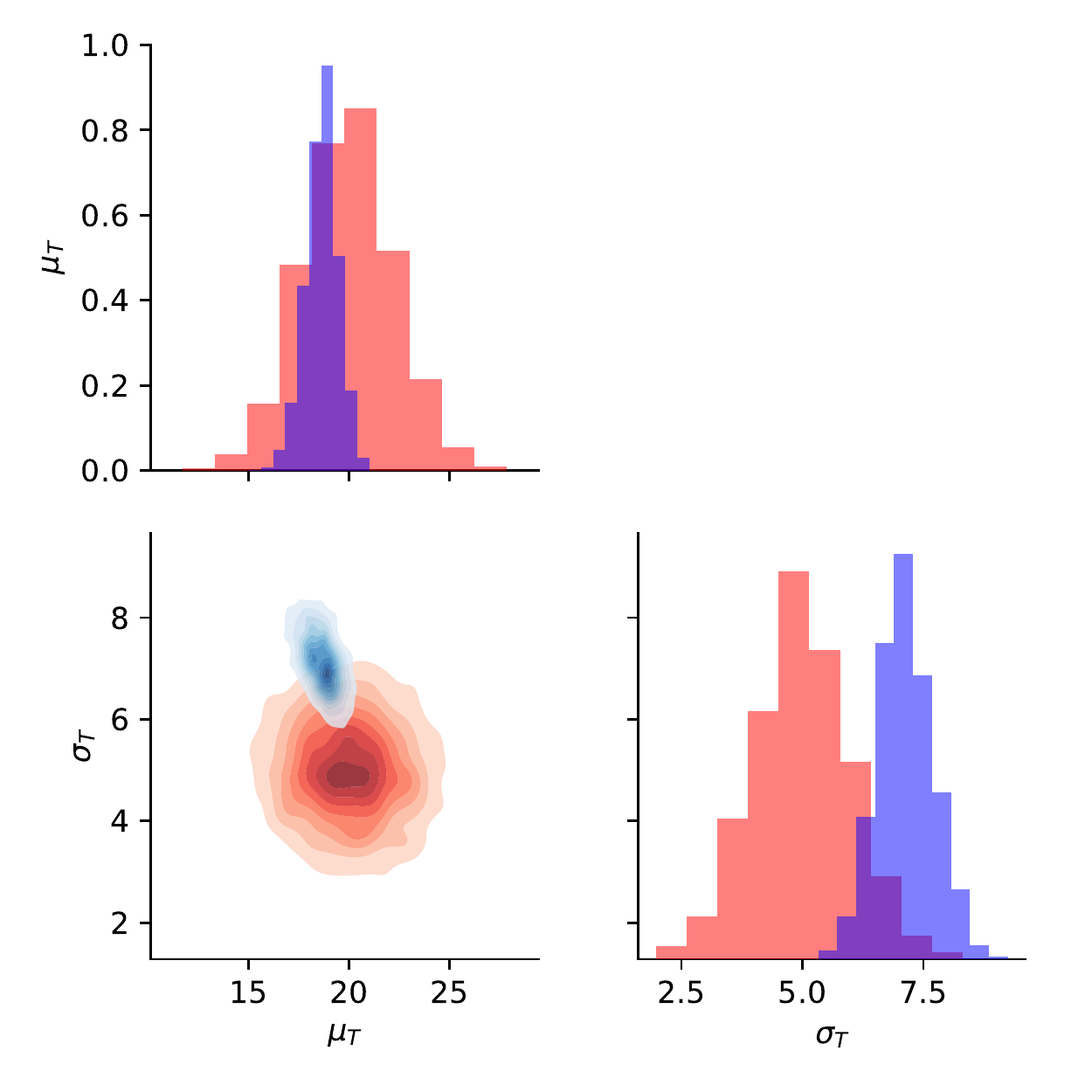}
\caption{\label{fig:2d_Gauss_prior_vs_post_temp_mean_vs_sigma} A comparison between the prior (red) and the posterior (blue) on the mean and sigma (width) values for the temperature component of the bivariate Gaussian model. This shows how the data constrains the parameter values.}
\end{figure}

\begin{figure}
\centering
\includegraphics[width=0.8\textwidth]{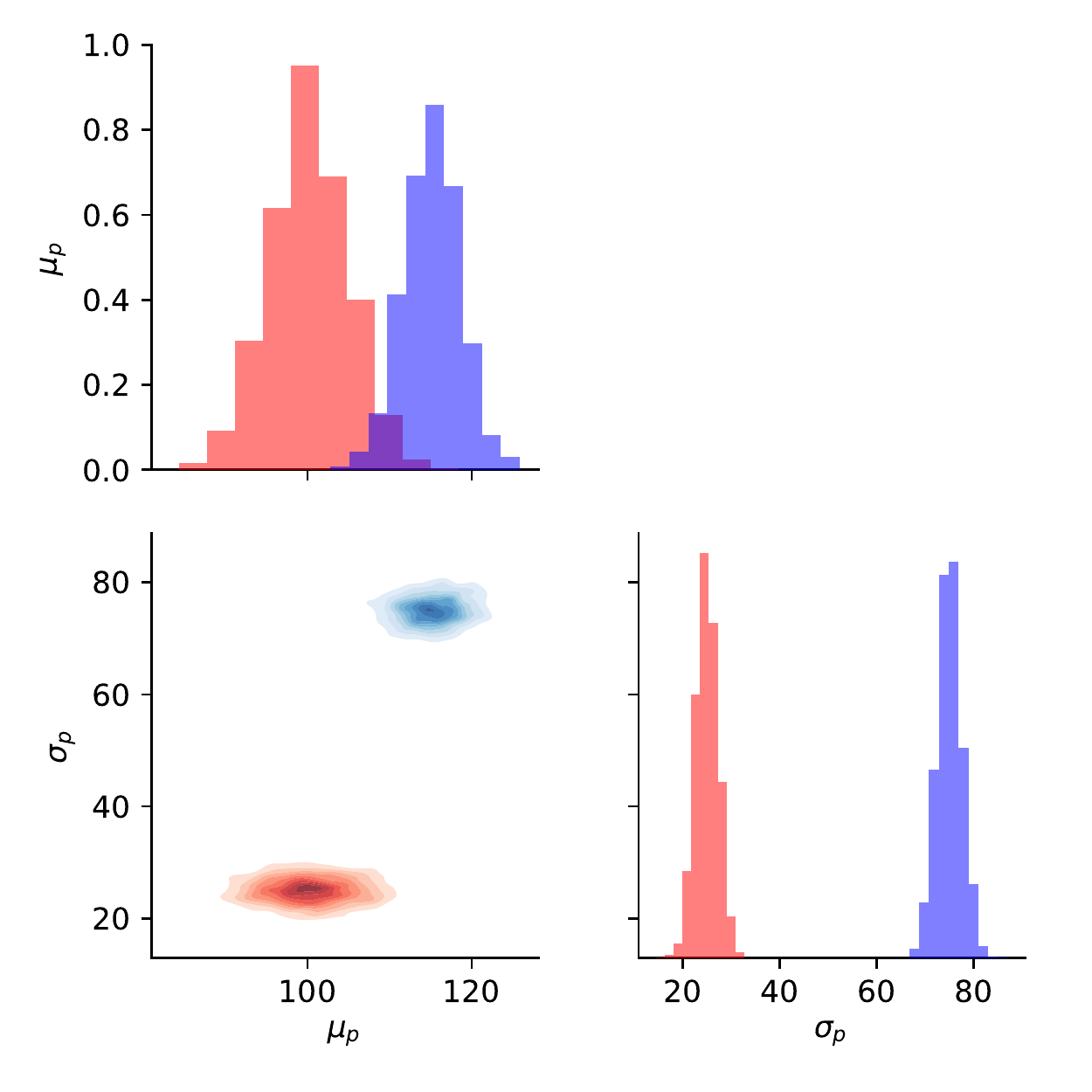}
\caption{\label{fig:2d_Gauss_prior_vs_post_precip_mean_vs_sigma} A comparison between the prior (red) and the posterior (blue) on the mean and sigma (width) values for the precipitation component of the bivariate Gaussian model. A large difference between the prior and posterior shows that the posterior is dominated by the likelihood and not the prior. This shows how the data constrains the parameter values.}
\end{figure}

\begin{figure}
\centering
\includegraphics[width=0.8\textwidth]{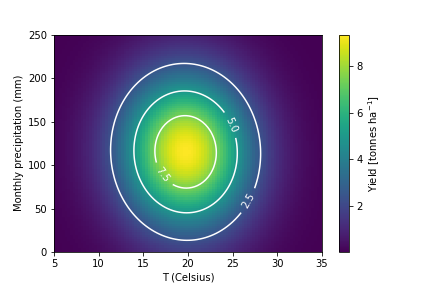}
\caption{\label{fig:response_function} Two-dimensional growth response function showing yield normalised to 2007 linear trend level as a function of monthly mean temperature and precipitation.}
\end{figure}

In this section we discuss the inferred parameter values of the bivariate Gaussian generative model and summarise its general form. Results are plotted in figures~\ref{fig:2d_Gauss_temp_post_sample_growth_curve}-\ref{fig:response_function}. These figures show the posteriors on the parameters of the bivariate Gaussian generative model. Figure~\ref{fig:response_function} shows the growth response function for the mean posterior parameter values. The mean parameter values for the posterior are $\mu_T = 19.1$~$^\circ$C, $\sigma_T = 6$~$^\circ$C, $\mu_P = 114$~mm, and $\sigma_P = 75$~mm. Statistics describing the posterior are presented in table~\ref{table:posterior}. The Gaussian formulation means that modelled growth will fall to less than 10\% of its maximal value for monthly temperatures that differ from the optimum ($\mu_T = 19.1$~$^\circ$C) by more than roughly two standard deviations, i.e. 2$\sigma_T = 12$~$^\circ$C. This two-sigma approach can be used to infer information about the cardinal temperatures for US maize \citep[e.g.][]{yin:1995}, and corresponds to lower and upper monthly temperature thresholds of roughly 7~$^\circ$C and 31~$^\circ$C, respectively.

\begin{table*}
\centering
\caption{Summary of the posterior on the parameters of the bivariate Gaussian. For each parameter of the model we present the posterior mean, the standard error on the mean, the standard deviation, the 2.5\%, 25\%, 50\%, 75\%, and 97.5\% percentiles the effective number of samples, $n_{\rm eff}$ and $\hat{R}$. $\hat{R}$ close to 1 is considered to be `good' and indicates convergence. There are 2000 samples from the posterior, $n_{\rm eff}$ takes correlations between chains into account so that it gives the equivalent number of completely independent chains. The correlation, $\rho$ (defined in equation~\ref{eqn:2d_gaussian_sigma}), is consistent with zero, indicating no evidence for a correlated response to temperature and precipitation. The normalisation sets the mean yield at actual T and P values equal to our baseline yield. For a full description of the model parameters see section~\ref{sec:2d_descrip_appendix}.}
\label{table:posterior}
\vskip 0.15in
\small
\begin{tabular}{lllllllllll}
Parameter             &  mean  & se$_{mean}$ & sd   & 2.5\%  & 25\%   & 50\%   & 75\%    & 97.5\% & $n_{\rm eff}$ & $\hat{R}$\\
\hline
$\mu_T$ [$^\circ$C]     & 19.14   &   0.02    &   0.65  &  17.73  & 18.73   &  19.17  &   19.59 & 20.28   &    1227 &    1.0\\
$\sigma_T$ [$^\circ$C]  & 6.22    &  0.02     &  0.51   &   5.3   &   5.87  &    6.2  &  6.53   &  7.35   &   918   &    1.0\\
$\mu_P$ [mm]            & 113.7   &   0.08    &   3.42  & 107.09  & 111.37  & 113.75  & 116.08  &  120.25 &  1872   &    1.0\\
$\sigma_P$ [mm]         & 75.47   &   0.06    &  2.57   &  70.57  &  73.73  &  75.42  &  77.22  &  80.55  &  2000   &    1.0\\
$\rho$                  & -0.03   &  1.5e-3   &  0.06   &  -0.14  &  -0.06  &  -0.03  &   0.01  &   0.09  &   1455  &    1.0\\
norm  [t/ha]            & 2.33    &  2.0e-3   &   0.06  &   2.21  &   2.28  &   2.33  &   2.37  &   2.45  &   1003  &    1.0\\
\end{tabular}
\end{table*}

Although they have been derived using monthly mean temperature and precipitation, the growth response functions can be expressed in terms of mean daily maximum temperature, giving a more direct comparison with previous work \citep[e.g.][]{schlenker:2009, hatfield:2011, sanchez:2014, hatfield:2015}. This makes use of a strong linear relationship between mean daily maximum and monthly mean temperature (Pearson correlation 0.998) as shown in figure \ref{fig:daily_vs_monthly}. One way of identifying the relationship between these related variables would be to fit their joint distribution. Here, we instead explore the conditional dependence by binning the monthly mean temperatures and calculating the associated mean daily maximum. The gradient of the best-fit line is 0.967 $\pm$ 0.024, and the intercept is 6.863 $\pm$ 0.34 $^\circ$C; consequently, an optimal monthly mean temperature of $\approx$ 18--19 $^\circ$C corresponds to an optimal daily maximum of $\approx$ 24--25 $^\circ$C. This is consistent with \cite{schlenker:2009} who applied a range of statistical models to US maize production and found increasing yield for daily temperatures up to 29$^\circ$C, and yield decreases above this threshold. The review by \cite{sanchez:2014} gives an optimal growing temperature of 30.8 $\pm$ 1.6 $^\circ$C for the whole maize plant life-cycle. However, the grain-filling stage of the cycle (typically June and July in the US) appears to be the most sensitive to temperature, with the optimal growing temperature given as 26.4 $\pm$ 2.1 $^\circ$C, overlapping with the results presented here. For completeness, the inferred monthly minimum and maximum temperature thresholds correspond to daily maxima of $\approx$ 13--14 $^\circ$C and $\approx$ 36--37 $^\circ$C, respectively. The symmetry of these values either side the optimal temperature is a consequence of the Gaussian formulation of the model, and differs to other estimates \citep[e.g.][and references therein]{zhou:2018}, but is broadly compatible. While this symmetry represents a limitation of the model, the Gaussian approach is still considered informative since it allows us to explore the joint non-linear response to temperature and precipitation.

As described earlier, \cite{kent:2017} investigated temperature and rainfall thresholds during June, July and August that were associated with large (>10\%) yield reductions, known as shocks. In that work, yield shocks were found to be associated with the simultaneous occurrence of mean temperatures above 23~$^\circ$C for the three month period, and total precipitation below 240 mm. Figure~\ref{fig:2d_Gauss_temp_post_sample_growth_curve} shows that 23~$^\circ$C is significantly greater than the optimal growing temperature, suggesting consistency with \cite{kent:2017}, despite applying very different approaches to the same data. Applying the offset between monthly mean temperatures and the daily maximum suggests that shocks are associated with extended periods during the grain filling stage with daily maximum temperatures above 29 $^\circ$C, showing excellent agreement with \cite{schlenker:2009}, and consistency with \cite{sanchez:2014} and \cite{hatfield:2015}. The 3-month precipitation threshold of 240 mm corresponds to an average of $\approx$~80 mm per month. This is significantly below the optimal monthly precipitation, shown in figure \ref{fig:2d_Gauss_precip_post_sample_growth_curve}, again indicating consistency between the two approaches. Within the Gaussian model described here, the thresholds identified by \cite{kent:2017} correspond to a 25\% reduction in yield relative to what would be expected under optimal temperature and precipitation conditions. Figures~\ref{fig:2d_Gauss_prior_vs_post_temp_mean_vs_sigma} and 
\ref{fig:2d_Gauss_prior_vs_post_precip_mean_vs_sigma} show the prior and posterior distributions of mean and variance for temperature and precipitation respectively and show how the posteriors are not sensitive to the assumed wide Gaussian priors.

Figure~\ref{fig:response_function} shows the global form of the growth response function, illustrating the joint response to both monthly mean temperature and precipitation. Based on this analysis, there is no evidence for correlated responses to temperature and precipitation; however, an absence of data in parts of the monthly T-P plane means that there is insufficient evidence to say that there is no correlated response. Had there been a strong correlation, maximising growth rates at higher temperatures would likely require higher precipitation, as for the emulator of rainfed mid-latitudes C4 crops, developed by \cite{snyder:2018}.



\section{Potential impacts of climate change on mean yield}
\label{sec:impacts}

The shape of the growth response function indicates the likely direction of climate impacts on maize yield in the US. In particular, as mean temperature increases, growing conditions will be further from the optimal for current varieties. This will tend to reduce yield in the absence of any successful adaptation strategies such as development of new maize varieties that are better suited to higher temperatures. An important caveat here is that we do not account for changing CO$_{2}$ concentrations, which are an input in the photosynthesis process. However, while increasing CO$_{2}$ levels might offset yield declines resulting from warmer temperatures, there is evidence that the photosynthetic rate for C4 crops starts to level out around 400 ppm \citep[e.g.][]{leakey:2009}.

\begin{figure}
\centering
\includegraphics[width=0.8\textwidth]{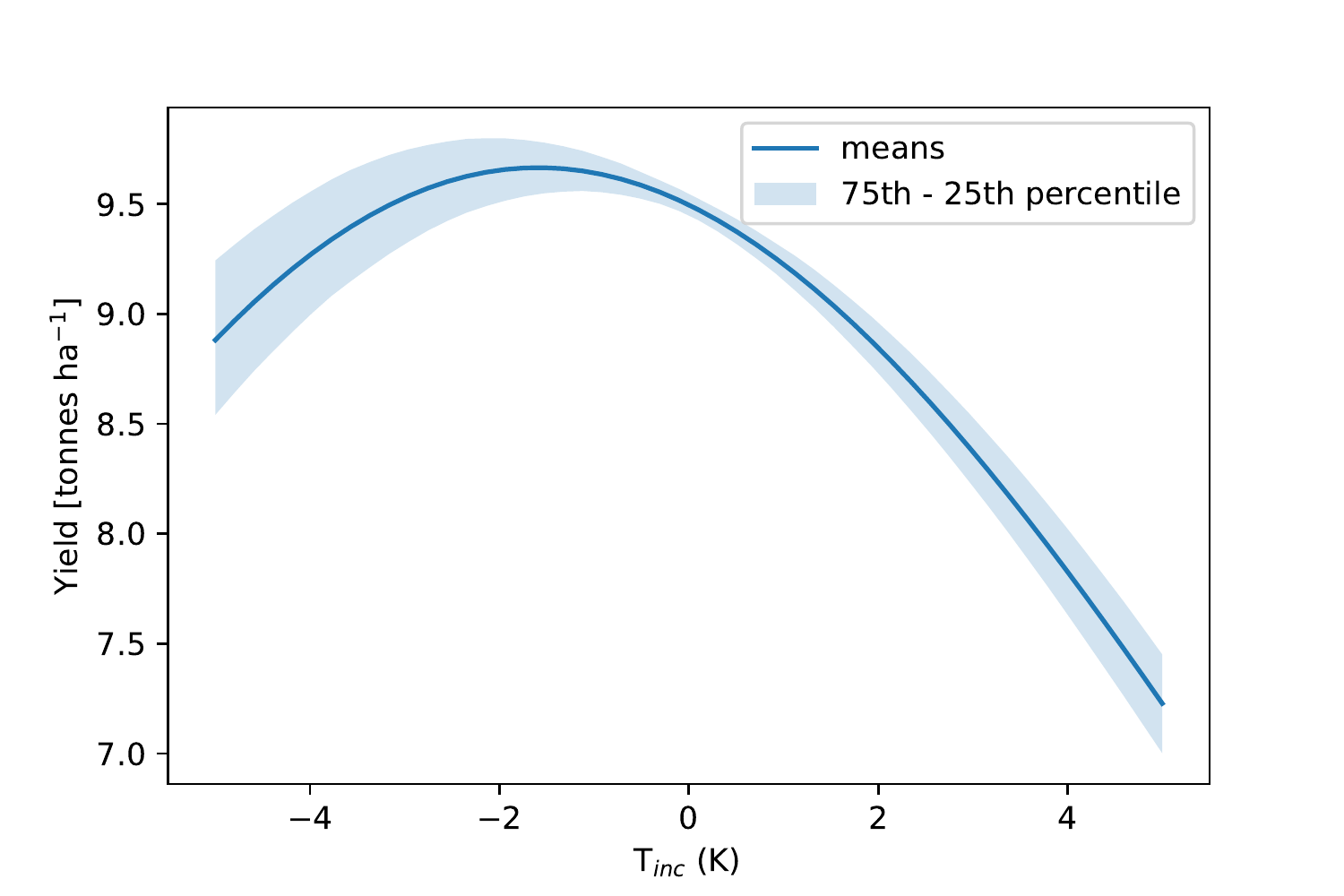}
\caption{\label{fig:yield_responses} Mean yield, normalised to 2007 linear trend levels, as a function of mean temperature change. The model predicts relative reductions in yield of $\approx 4$\% for a mean temperature increase of 1~$^\circ$C. }
\end{figure}

To test the climate change response of the model, we used the derived growth response function to quantify the mean yield as a function of changes in mean temperature, while keeping precipitation and growing area fixed. To ensure a fair comparison with the present-day, we applied the delta change method by adding a constant increment to the observed temperature time series, and modelling the resultant yield. The temperature increment is varied between -5 and 5$^\circ$C  as shown in Figure \ref{fig:yield_responses}. In agreement with previous studies \citep[e.g.][]{bassu:2014, urban:2015, lobell:2017}, this suggests that the US maize yield is likely to decrease by several percent in response to a 1 $^\circ$C temperature rise, with larger reductions expected for greater warming.

\begin{figure}
\centering
\includegraphics[width=0.8\textwidth]{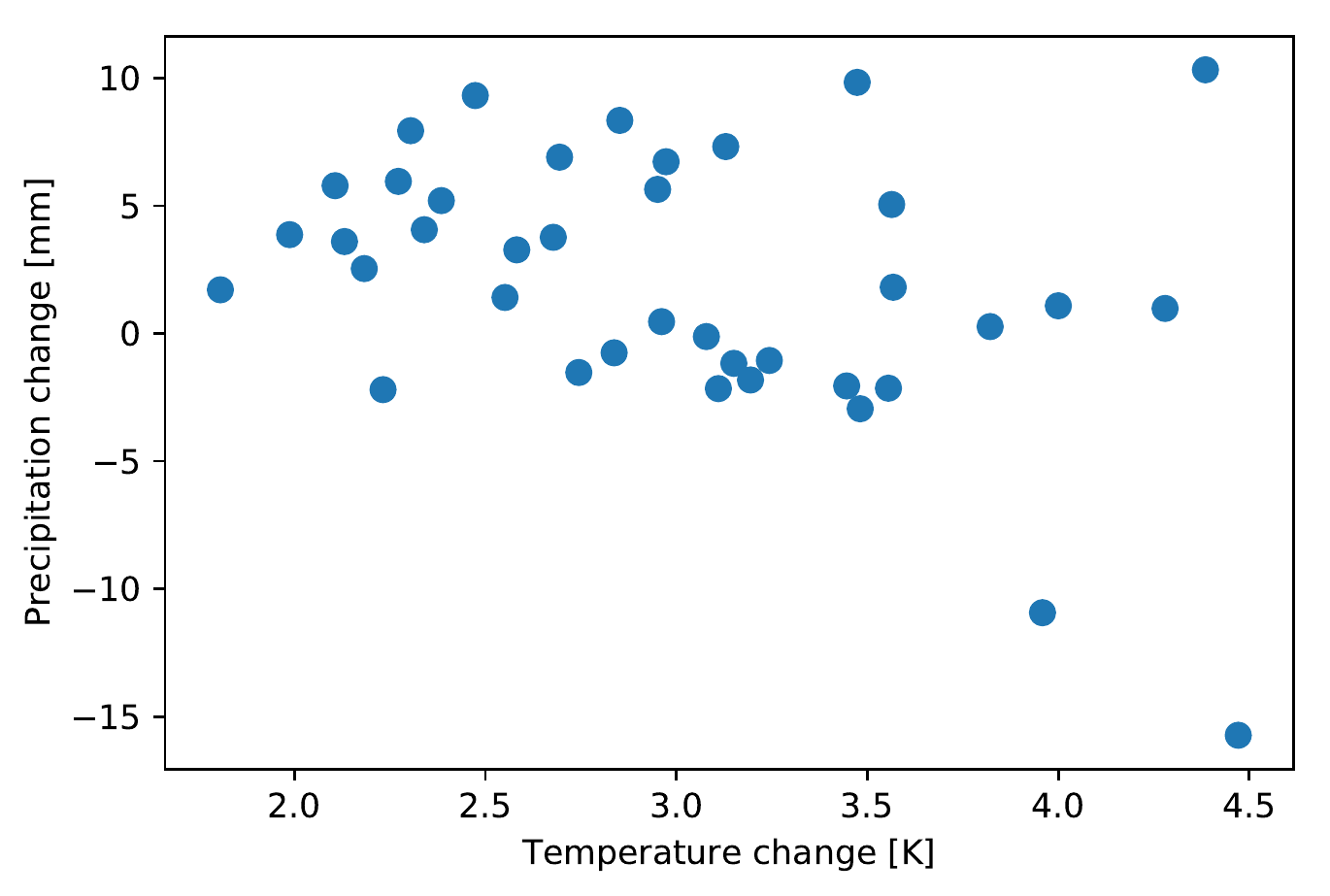}
\caption{\label{fig:sim_p_t_dist} Simulated changes in temperature and precipitation between 2041-2070 relative to 1981-2014 under the RCP8.5 scenario, \citep[][]{ipcc2:2013}. }
\end{figure}

Figure~\ref{fig:yield_responses} provides a more complete exploration of potential climate change impacts within this framework, showing the expected yield as a function of constant changes in both temperature and precipitation, applying independent increments of -5 to 5 $^\circ$C and -100 to 100 mm to the observed temperature and precipitation time series, respectively. The precipitation increments are broadly in line with projected changes in the climate during the 21st century \citep{usgcrp:2017}, see also Figure~\ref{fig:sim_p_t_dist}. 

\begin{figure}
\centering
\includegraphics[width=0.8\textwidth]{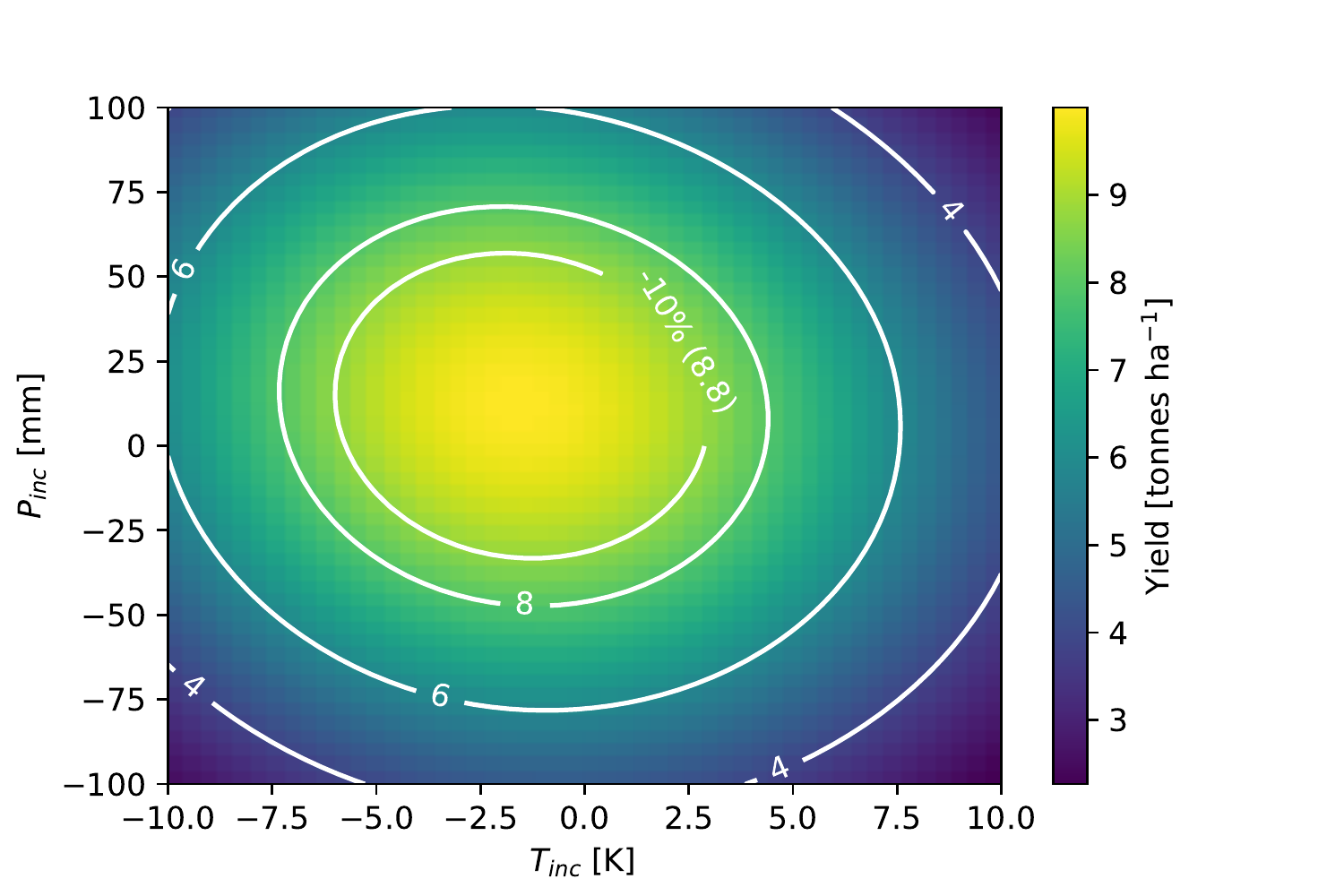}
\caption{\label{fig:temp_precip_impact} Yield as a function of mean temperature and precipitation change normalised to 2007 linear trend levels. The model predicts reductions in yield due to increasing temperatures and decreasing precipitation. We include a contour at a yield of 10\% below the current value to define a rough region outside which we see fractionally large reductions.}
\end{figure}

This provides a look-up table of yield as a function of simultaneous changes in temperature and precipitation, subject to the caveats outlined previously. For example, figure~\ref{fig:temp_precip_impact} suggests that US maize is relatively well-suited to the current climate and that significant changes in either/both temperature and rainfall are likely to reduce mean yield. In contrast, locations where current temperatures are below the optimum (e.g. the UK), would be expected to see an increase in maize yield with a warming climate. Due to the formulation of the model, this analysis does not allow for changes in the frequency of extreme daily maximum temperatures or precipitation intensity which may have a significant influence on both average yield and yield variability.

\begin{figure}
\centering
\includegraphics[width=0.8\textwidth]{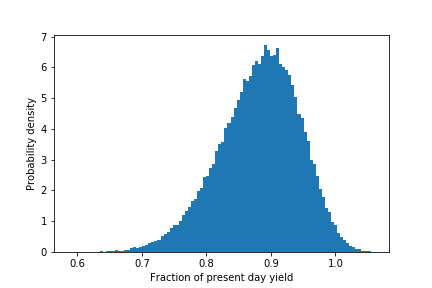}
\caption{\label{fig:simulated_impact} Influence of simulated temperature and precipitation changes \citep[][]{ipcc2:2013} on mean US Maize yield in 2041-2070 relative to 1981-2014 under the RCP8.5 scenario, keeping growing area fixed. The distribution is determined by the simulated changes in temperature and precipitation, and the posterior of the fitted model - we find a median reduction of 12\%. }
\end{figure}

Finally, to demonstrate the flexibility of the model, we have estimated yield based on projected changes in temperature and precipitation calculated by global climate model simulations provided by the Intergovernmental Panel on Climate Change Fifth Assessment Report, Working Group 1 \citep[http://www.climatechange2013.org/report/full-report/][]{ipcc2:2013}. Figure~\ref{fig:sim_p_t_dist} shows the ensemble of projected changes in summer temperature and precipitation for the Central North American Giorgi region (which encompasses the US Cornbelt) under the RCP8.5 scenario for 2041-2070 relative to 1981-2014. For each of the 39 climate model simulations, the projected change in temperature and precipitation is added to the historical climate data to create synthetic time series of future weather conditions. We then compute the yield for the full sample of 2000 parameter values from the posterior, giving a probability distribution of yields with a median reduction of 12\%. Figure~\ref{fig:simulated_impact} shows the full distribution of yield changes arising from these changes using the yield model presented here.

\section{Discussion}
\label{sec:future}
Understanding climate change impacts on food production is essential for developing effective policies and adaptation plans at local, national and international scales that will ensure food security for all. The broad aim of the work presented here is to explore the feasibility of deriving physically-realistic relationships between maize yield in the US and local monthly temperature and precipitation during the growing season. A key component of this was developing a computationally inexpensive generative model that captures the main impacts of monthly meteorology on maize growth rates, using this understanding to assess the plausible impacts of climate change on current varieties in the absence of adaptation measures. In turn, this new approach can contribute to the growing body of evidence that supports genetic breeding programs and the development of improved agronomic practices which will ensure high levels of agricultural productivity in the future. One caveat is that this data-driven approach is not designed to capture all of the relevant physiological processes that govern maize growth and will fail to identify high impact events associated with the plant's response to pests, disease, pollution or short-lived meteorological extremes (e.g. frost, hail, high winds, extreme rainfall, flash droughts, etc).

The reliability of the empirical growth response functions is dependent on the quality of both the yield and meteorological data. These data are well-documented in the US, making it an ideal location for exploring and validating the approach, i.e. comparing critical temperatures with laboratory measurements and other studies \citep[e.g.][]{cutforth:1990, schlenker:2009, hatfield:2011, lobell:2013, sanchez:2014}. In other regions, historical records may be sparser, potentially making the modelling more challenging; however, we have demonstrated here that it is possible to extract useful information about crop response functions from monthly data at only a few different locations. For that reason, this approach lends itself to locations where data is limited, and it will be of interest to assess how well the approach can capture the characteristics of different crops grown in different environments.

The model presented here has been developed firstly as a way of capturing the broad influence of natural climate variability on US maize yield, and secondly to explore potential climate change impacts on current maize varieties in the absence of adaptation. Because of this, our model is closely related to previous work \citep[e.g.][]{schlenker:2009, roberts:2012, lobell:2013}, while our model also provides a method for the robust characterisation of model errors and permitting applications with differing time resolution measurements. We have also demonstrated how, given projections of monthly temperature and precipitation from global climate simulations, our model can be used to provide a probability distribution of the impact on mean yield.

We emphasise that the model demonstrated here only considers area-average yield, rather than total production which may be more relevant for assessing potential climate change stresses on national food security. The reason for this is that total production is a function of both yield and the area harvested. The latter is known to be affected by a range of non-climate factors, including commodity and crude oil prices \citep[e.g.][]{Zafeiriou:2018}, as well as water availability and soil suitability. In contrast, the yield is expected to be much more strongly correlated with weather conditions during the growing period.

Previous research has explored the influence of a range of adaptation strategies \citep[e.g.][and references therein]{challinor:2014}. Examples include changes in crop varieties, species, planting times, irrigation, as well as more transformational changes such as crop relocation. Other studies have focused more on the economic implications of adaptation \citep{Seo:2008, schlenker:2013, Carter:2018, Dalhaus:2018}. In principle, models can also incorporate a number of adaptation options, such as different crop varieties or species, planting dates and irrigation. 

The approach presented here contributes to the toolkit of methods that seek to inform adaptation decisions, particularly phenotyping and the objectives of crop genetic improvement programmes (e.g. http://www.wgin.org.uk/) such as helping crop breeders identify and target traits that will be more beneficial in the future (e.g. heat and drought tolerance, higher optimal growing temperatures), as well as assisting agronomists in developing improved practices that will maintain high levels of productivity. The results can also provide useful context for interpreting climate change impacts simulated by more complex models \citep[e.g.][]{tigchelaar:2018, ostberg:2018}.

More generally, this approach could be used to identify regions where current crop varieties and agronomic practices are projected to come under stress in the future, indicating where and when incremental or transformational adaptation may be most effective.

\section{Conclusions}
\label{sec:conclusions}
We have developed and applied data-driven statistical models for exploring the dependence of US maize yield variations on monthly mean temperature and precipitation, using both linear and non-linear relationships. However, a particular aim of the approach presented here has also been to assess the feasibility of extracting physically plausible growth rate information from limited data, in this case state-scale yield and monthly meteorological information. We have demonstrated how these coarser grained models can be used with predicted global meteorological changes to compute yield reduction risks.

The linear model predicts maize yield using a sum over the growing season of the monthly contributions to growth based on a linear function of observed temperature and precipitation. The early months in a calendar year are weak predictors of maize yield, consistent with a planting date around April, c.f. AMIS crop calendar. In contrast, the following six months to harvest (i.e. April-September) are found to be strong predictors of maize yields, indicating that weather conditions during the growing season in the US Cornbelt are statistically much more important than antecedent conditions.

The non-linear model predicts maize yield using a sum over the growing season of the monthly contributions to growth based on a time-stationary two-dimensional Gaussian function of observed temperature and precipitation. The modelled growth rates are maximal at the peak of the function, and lower either side \citep[c.f.][]{cutforth:1990, wang:1998,  streck:2007, hatfield:2015, korres:2016, tigchelaar:2018}. As such, the growth response function represents the typical response of the crop, averaged over the growing season. The yield and meteorological data are then used to constrain the location of the peak and the width of the bivariate Gaussian function, which provides information about the optimal monthly temperature and rainfall for current US maize varieties.

There are several major differences between this approach and previous work: firstly, our model incorporates non-linear responses to both temperature and precipitation, allowing for potential interactions to impact maize yield; secondly, the non-linear function is Gaussian rather than a Beta function \citep[e.g.][]{yin:1995, streck:2007}, which avoids the need to estimate explicit minimum and maximum thresholds for temperature and precipitation since the growth rate tend to zero for large deviations from optimal growing conditions; thirdly, the contributions to maize growth are calculated and summed on monthly intervals, rather than daily as for Growing Degree Days \citep[e.g.][]{zhou:2018}.

This approach represents a simplification of the crop's true response function which is time-dependent and multivariate \citep[e.g.][]{siebert:2017}, reflecting changes in the crop's sensitivity to meteorological conditions during different growth phases \citep[e.g.][]{hatfield:2011, sanchez:2014}. Despite this, the formulation shares similarities with other models that have explored non-linear yield responses to temperature and accumulated thermal units \citep[e.g.][]{lobell:2013, zhou:2018}. In addition, it also allow us to straightforwardly explore the joint influence of temperature and rainfall variations on monthly timescales. Within this framework, we find maize growth rates are maximal for a monthly mean temperature of $19 \pm 0.7 ^\circ$C, and monthly total precipitation of $ 114 \pm 3$ mm. This corresponds to a daily maximum temperature of 24--25~$^\circ$C, in approximate agreement with \cite{sanchez:2014, hatfield:2015} for the grain filling phase of growth. Due to the shape and fitted parameters of the bivariate Gaussian function the growth rates decline rapidly for temperatures above this threshold, in agreement with \cite{schlenker:2009}. 

Our analysis also suggests that current US maize varieties are relatively well optimised for present-day growing conditions in the US Cornbelt, but that growth rates would be maximised at slightly lower monthly temperatures ($\approx$ -1.5 $^\circ$C) and slightly higher monthly precipitation totals ($\approx$ +25 mm). Keeping precipitation at present-day levels and the growing area fixed, a 1$^\circ$C temperature rise is expected to reduce the mean yield by 3-5 \%, in broad agreement with previous findings \citep[e.g.][]{bassu:2014, urban:2015, lobell:2017}. However, we note that this analysis does not allow for various adaptation strategies (such as changes in planting date, location or irrigation) or account for changes in carbon dioxide (which may not have a strong influence at current concentrations \citep[e.g.][]{leakey:2009}). Similarly, changes in the frequency of extreme daily maximum temperatures or precipitation intensity may have a significant influence on both average yield and year-to-year variability. 

The similarity with previous findings is encouraging given that, to our knowledge, this is the first attempt to directly estimate a growth response function from state-scale data, rather than from field trials. This suggests that the new approach could be applied across a range of spatial and temporal scales, and to distinguish growth responses of different maize varieties. The main constraint to this application is the need for sufficient yield and meteorological data to robustly estimate the model parameters outlined in table~\ref{table:posterior}.  

More broadly, this approach provides an intuitive and computationally inexpensive method for deriving data-driven crop indices that can be used in climate risk studies \cite[e.g.][]{kent:2017}, and can complement other approaches to modelling crops \citep[e.g.][]{schlenker:2009, lobell:2010, carter:2016, zhou:2018}. 

\section*{Acknowledgements}

This work was undertaken as the Forecasting Agricultural Crop Yields on National scales (FACYNation) project, supported by funding from the Science and Technology Facilities Council (STFC) through the STFC Food Network+. Raphael Shirley acknowledges support from the Daphne Jackson Trust and from the Spanish Ministerio de Ciencia, Innovaci\'{o}n y Universidades (MICINN) under grant numbers ESP2017-86582-C4-2-R and ESP2015-65597-C4-4-R. Edward Pope, Chris Kent and James Bacon were also supported by the Met Office Hadley Centre Climate Programme (GA01101) and thank Stewart Jennings for useful discussions. We are extremely grateful to the anonymous reviewers for their help improving the work and its presentation.




\interlinepenalty=10000

\bibliographystyle{plainnat}
\bibliography{./bibliography}




\appendix


\section{Access to the code}
\label{sec:model_descrip_appendix}

All the code and data is available on GitHub:\\

\url{https://github.com/raphaelshirley/facynation}\\ 

All the models in the Stan format are there to be rerun or extended through Jupyter notebooks.


\label{lastpage}
\end{document}